# A Compact PCI-based Measurement and Control System for Satellite-Ground Quantum Communication

Binxiang Qi, Shubin Liu, Qi Shen, Shengkai Liao, Wenqi Cai, Zehong Lin, Weiyue Liu, Chengzhi Peng, and Qi An

***Abstract*—Since the 1990s, there has been a dramatic interest in quantum communication. Free-space quantum communication is being developed to ultra-long distance quantum experiment, which requires higher electronics performance, such as time measurement precision, data-transfer rate, and system integration density. As part of the ground station of quantum experiment satellite that will be launched in 2016, we specifically designed a compact PCI-based multi-channel electronics system with high time-resolution, high data-transfer-rate. The electronics performance of this system was tested. The time bin size is 23.9ps and the time precision root-mean-square (RMS) is less than 24ps for 16 channels. The dead time is 30ns. The data transfer rate to local computer is up to 35 MBps, and the count rate is up to 30M/s. The system has been proven to perform well and operate stably through a test of free space quantum key distribution (QKD) experiment.**

## I. Introduction

OVER the past two decades, quantum communication has become the hotspot of future information technology research because of its unconditional communication security which is guaranteed by the laws of quantum physics [1]. QKD, more popularly known as quantum cryptography, has been the most developed application of quantum communication [2]. An overview of QKD system is shown in Fig. 1. The Bennett-Brassard 1984(BB84) [3] protocol is the well-known provably secure QKD protocol, which provides completely secure key between communicating parties, the sender (typically named Alice) and the receiver (typically named Bob). Alice and Bob have both a quantum channel and a classical channel for communication. The quantum channel is only used to transmit single quanta (qubits) and must consist of a transparent optical path such as optical fiber and free space path. The classical channel can be a conventional internet protocol (IP) channel such as Ethernet connection, telephone line, and optical communications link, which transmits ciphertexts and authentication information. Sending randomly encoded information on single photons produces a shared secure key that is a random string, and the probabilistic nature of measuring the photon state provides the basis of its security. So an eavesdropper, commonly referred to as Eve, can’t make an undiscovered intercept due to the quantum mechanical properties of single photons. Any potential eavesdropping can be detected via the quantum bit error rates (QBER) of the quantum channel. In practice, errors introduced by imperfect equipment and environmental noise can mask the activities of Eve.

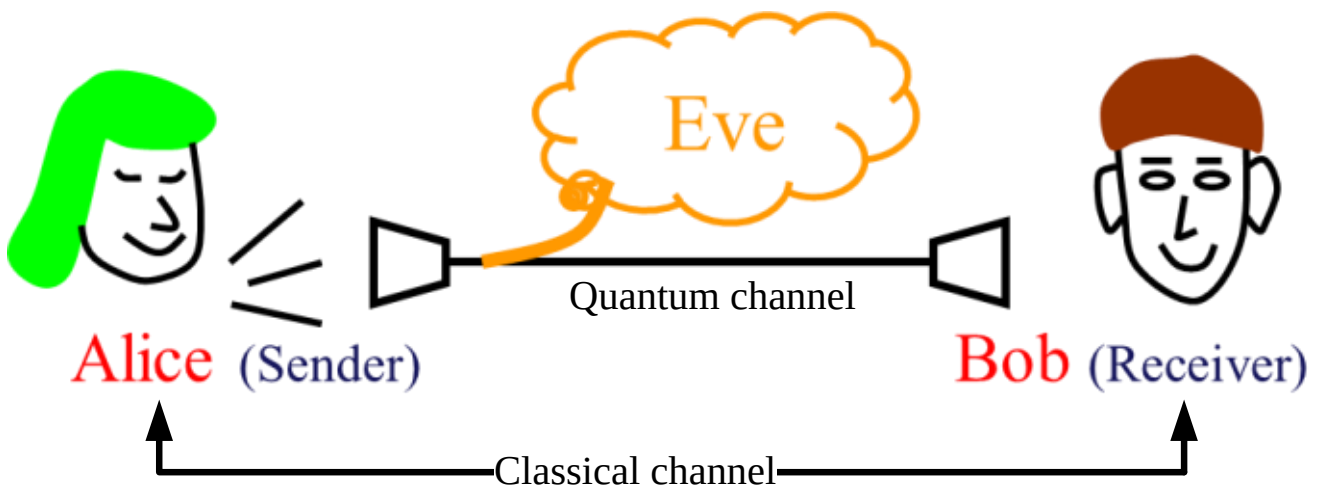

Fig. 1 Overview of QKD system. Where Alice as a sender, Bob as a receiver, Eve as a eavesdropper

Because of the low atmospheric absorption for certain ranges of wavelength, free-space links could have lower channels loss compared to the optical fiber links. Due to the effective thickness of the atmosphere is equivalent to 8-10 km of ground atmosphere[4], Free-space quantum communication has been demonstrated as the most feasible option to achieve ultra-long

Manuscript received June 16, 2014. The work was supported by the Chinese Academy of Sciences (CAS), the Fundamental Research Funds for the Central Universities under Grant No. WK2340000056, and the Zhejiang Provincial Natural Science Foundation of China under Grant No. LY13F050007.

Binxiang Qi, Shubin Liu and Qi An are with the State Key Laboratory of Particle Detection and Electronics, and Department of Modern Physics, University of Science and Technology of China, Hefei, 230026, China(e-mail: qibx@mail.ustc.edu.cn; liushb@ustc.edu.cn).

Qi Shen, Shengkai Liao, Chengzhi Peng and Wenqi Cai are with the Hefei National Laboratory for Physical Sciences at Microscale, and Department of Modern Physics, University of Science and Technology of China, Hefei, 230026, China (e-mail: shenqi@ ustc.edu.cn; skliao@ustc.edu.cn).

Binxiang Qi, Qi Shen, Shengkai Liao, Shubin Liu, Wenqi Cai, Chengzhi Peng and Qi An are with the Synergetic Innovation Center of Quantum Information and Quantum Physics, University of Science and Technology of China, Hefei, 230026, China

Zehong Lin and Weiyue Liu are with the College of Information Science and Engineering, Ningbo University, Ningbo, China.

distance quantum communication [5][6]. With a series of experimental researches and full-scale verifications [4][7][8], satellite-ground quantum communication projects have been proposed by several countries [9][10]. Being one of strategic space projects proposed by Chinese Academy of Sciences (CAS), quantum experiment satellite is being developed and will be launched in 2016, which aims for realization of satellite-ground quantum communication network and test of quantum foundations on a global scale.

In this paper, based on the compact peripheral component interconnect (CPCI) bus, we designed a measurement, control and data-transmission electronics system as part of the ground station of quantum experiment satellite.

## II. Requirements

In quantum communication system, secure key is transmitted by encoding it in the quantum state of a series of single photons. Fig. 2 [11] shows a typical free-space QKD system based on the BB84 protocol. Bob needs to detect and record the arrival times of the single photons sent by Alice, and they are often far away from each other and have independent reference clocks. For electronics, a time measurement module is essential to accurately record the arrival time of single photons and establish high-precision time synchronization between Alice and Bob. QKD control module is used to generate proportion-adjustable random code to encode information on single photons, assemble the experimental data and control the experiment. The experimental data is saved to local CPU via the interface. A GPS signal is used to synchronize clocks over large distance with high precision and accuracy.

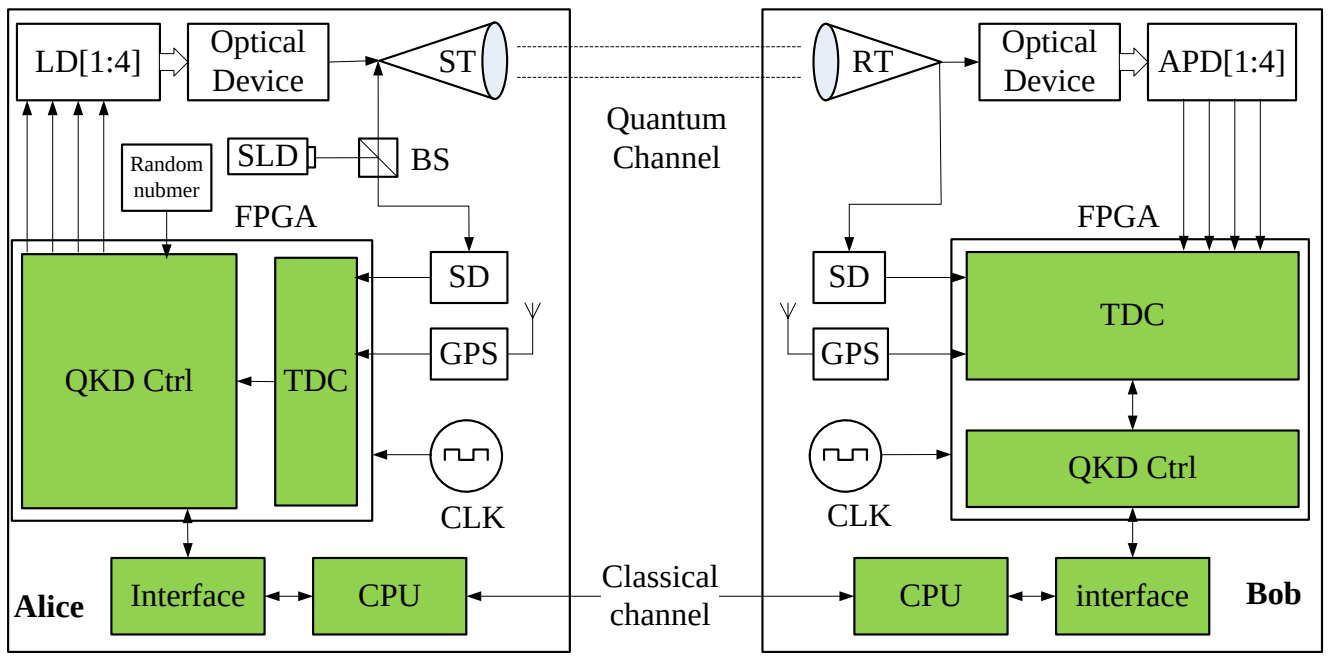

Fig. 2 A schematic of typical free-space QKD system

The timing resolution and the time synchronization precision can directly influence the performance of QKD systems, such as QBER, final key rate, etc. [13]. The high time precision can improves the system timing jitter, and then the time coincidence window width can be set smaller, and results in a better SNR with more noise eliminated. The better SNR causes smaller QBER.

In the satellite-ground quantum communication, the system requires that ground station electronics should have following performance. The time measurement is less than 30 ps. The random number generation rate is greater than 2 Mb/s and data-transfer-rate is greater than 20 MB/s. The quantum bit error rate (QBER) is less than 3.5% and the final key rate is greater than 500 bps. In addition, it should have the ability of photon event counting and system monitor to debug and test.

## III. Architecture of the Electronics

Based on the CPCI system, the electronics system is designed in a standard 6U size circuit board. Fig. 3 shows its architecture, which mainly consists of multi-channel input module, time measurement module, experiment control module, data transmission module, system monitor module and output module. Besides, a counter is designed for experimental debugging and verification test.

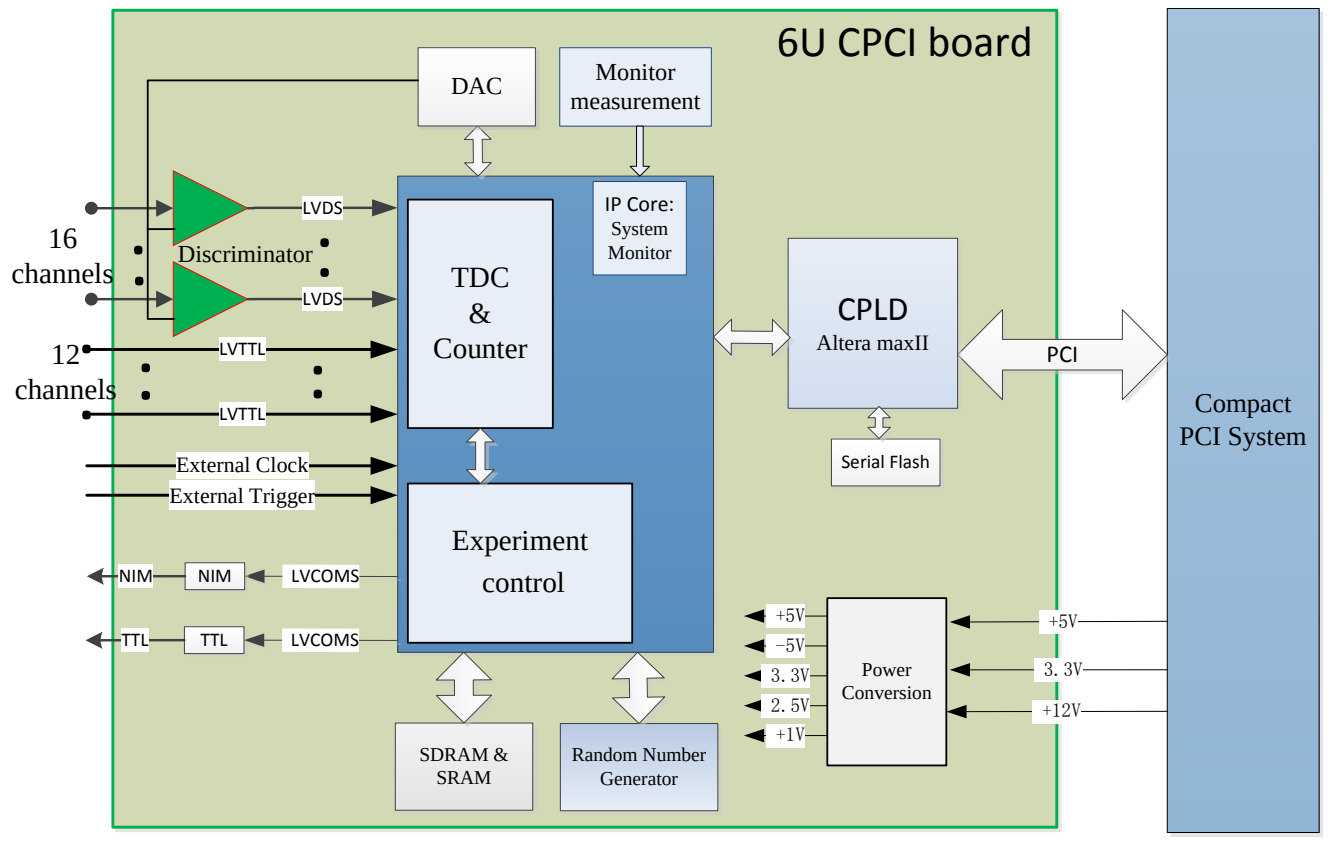

Fig. 3 Architecture of the electronics

### A. *Time Measurement*

A 16-channel high resolution time-to-digital converter (TDC) is implemented in a Xilinx Virtex6 field programmable gate array (FPGA) to record the arrival time of the input signals. The block diagram of TDC is shown in Fig. 4, which consists of the coarse measurement unit, the fine measurement unit, encoder unit and the readout buffer. The fine measurement unit is realized by time interpolation using the dedicated carry chains in the FPGA [11][12]. The coarse measurement unit is achieved by a simple counter with the system clock at 160 MHz. We use the half-interval search algorithm to convert thermometer code to binary code for encoder unit. Clock samples the state of delay cells when a hit arrives. By determining in which sample the rising edge of input signal comes out of a delay cell, the arrival fine time of the input signal can be deduced with a resolution equal to the tap delay. The system clock drives coarse counters to track the number of multiplied clock periods elapsed since the TDC was enabled. The coarse counter values are sampled when the hit arrives. A TDC data is assembled with coarse time data, fine time data and channel identification.

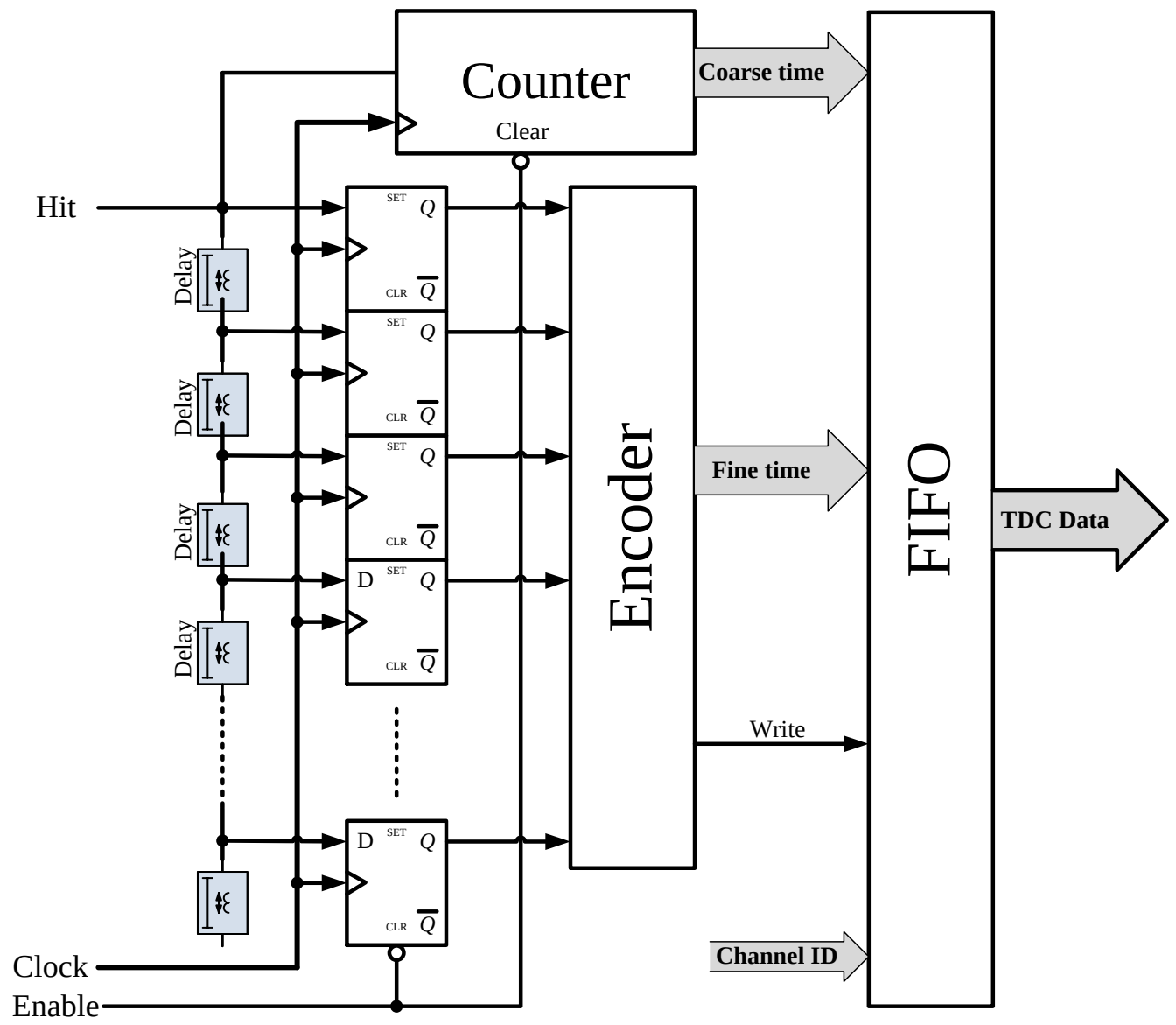

Fig. 4 Block diagram of TDC based on FPGA's carry chains

## B. Experiment Control

The core of the electronics system is experimental control, which controls the timing of the quantum communication experiment. Fig. 5 shows its block diagram and Fig. 6 shows its control flow chart. It has the following functions: parsing commands and transmitting them to other modules, maintaining the system absolute time by receiving global positioning system (GPS) signal or a local signal, generating proportion-adjustable random code and outputting the code with a fixed frequency, assembling the random number data and TDC data to the experimental data and then writing the data by frame format to buffer FIFO. A physical random generator provides enough random numbers by before the start of communication and stores them in a local memory. A synchronous dynamic random access memory (SDRAM) is used to guarantee no data loss during processing.

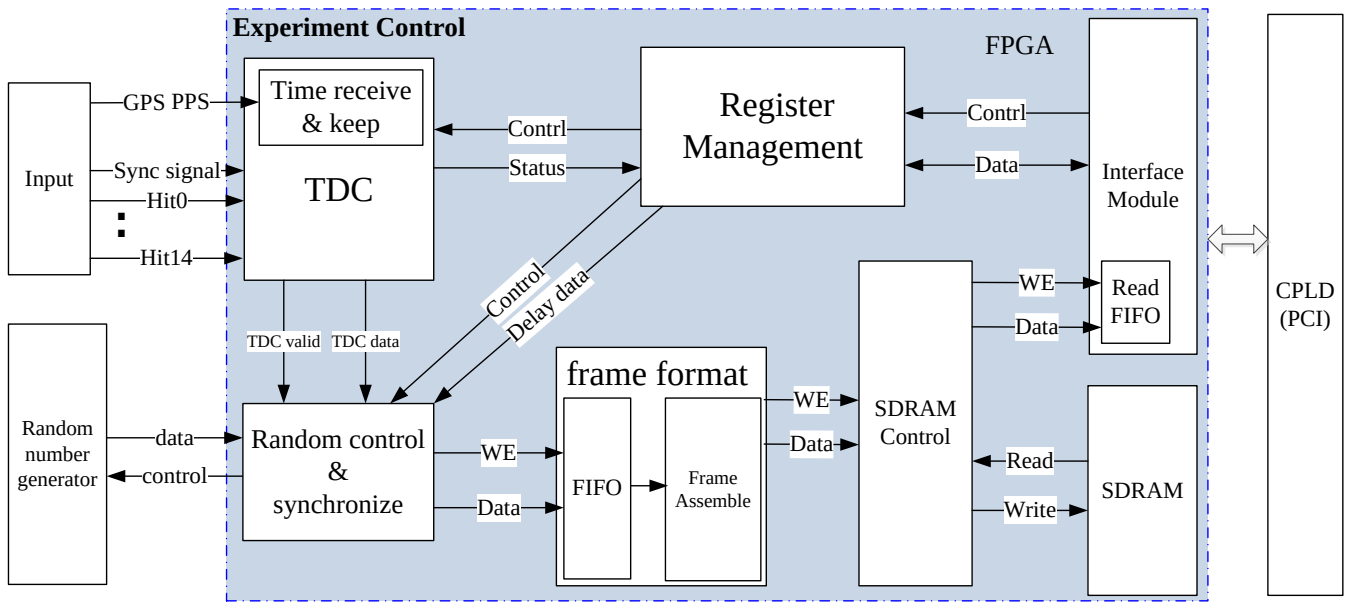

Fig. 5 Block diagram of Experiment Control based FPGA

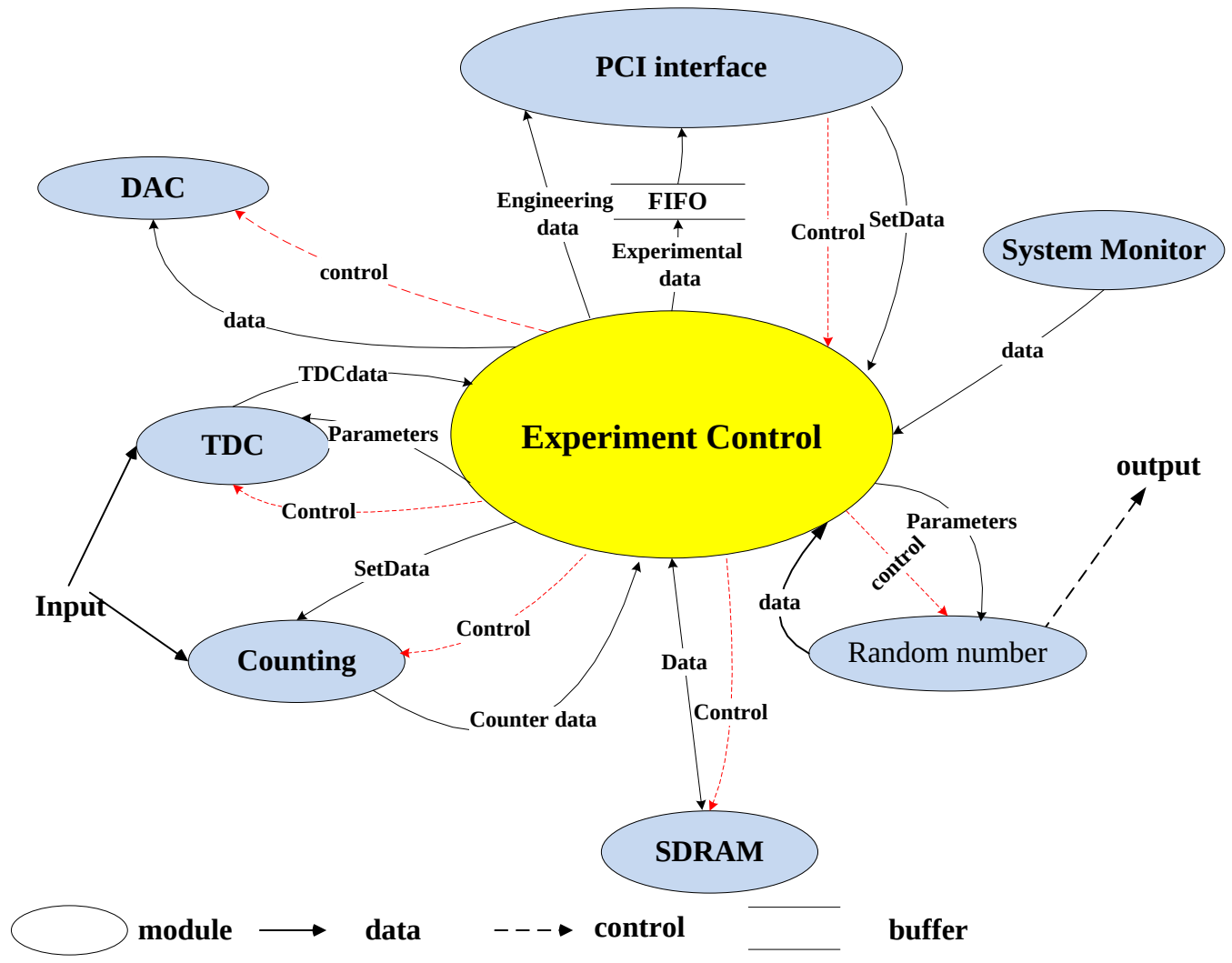

Fig. 6 Control flow chart of Experiment Control

## C. Data Transmission

The data transmission is based on PCI protocol to guarantee a high data rate. As shown in Fig. 7, a complex programmable logic device (CPLD) with the IP core pci_mt32 [14] functions as the interface between the FPGA and the PCI bus. The data of quantum communication experiment is timely transferred to the PC through the PCI interface. To achieve a high transfer speed, the DMA (Direct Memory Access) method is adopted. With different DMA transfer length, the average transfer speed varies.

The on-line modification of the FPGA logic is achieved with the passive serial (PS) configuration mode through this CPLD via PC. The CPLD controls the transfer of the configuration data from a storage device (a Sflash memory) to the FPGA. In the on-line modification process, the Sflash controller section erases the Sflash and reads new configuration data from the PCI bus, and then read data from the Sflash to configure the FPGA.

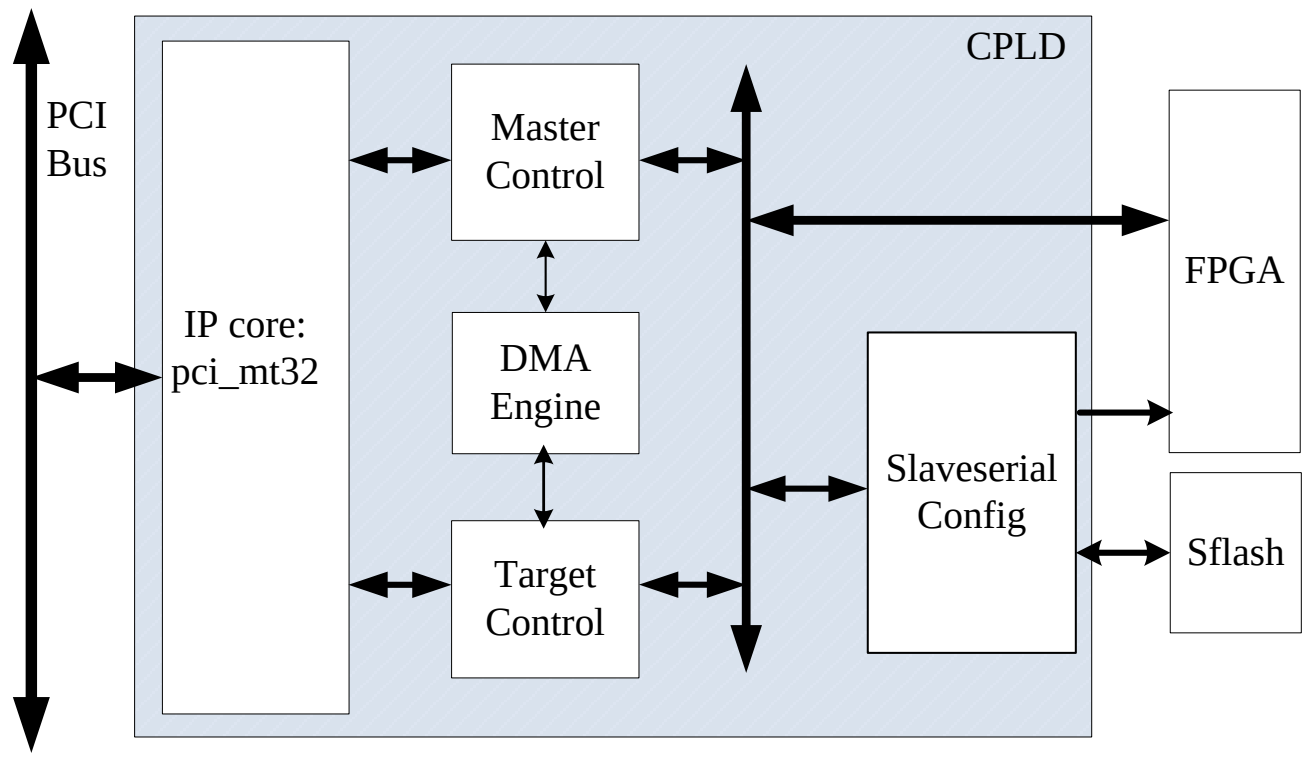

Fig. 7 Block diagram of data transmission based on CPLD

## D. The Others

As shown in Fig. 8, a system monitor module is used to measure the voltage and current on board, and FPGA's temperature and voltage. Sense resistor detected the flowing current, and the analog signal of voltage and current is sampled using IP core [15] provided by Xilinx Inc.

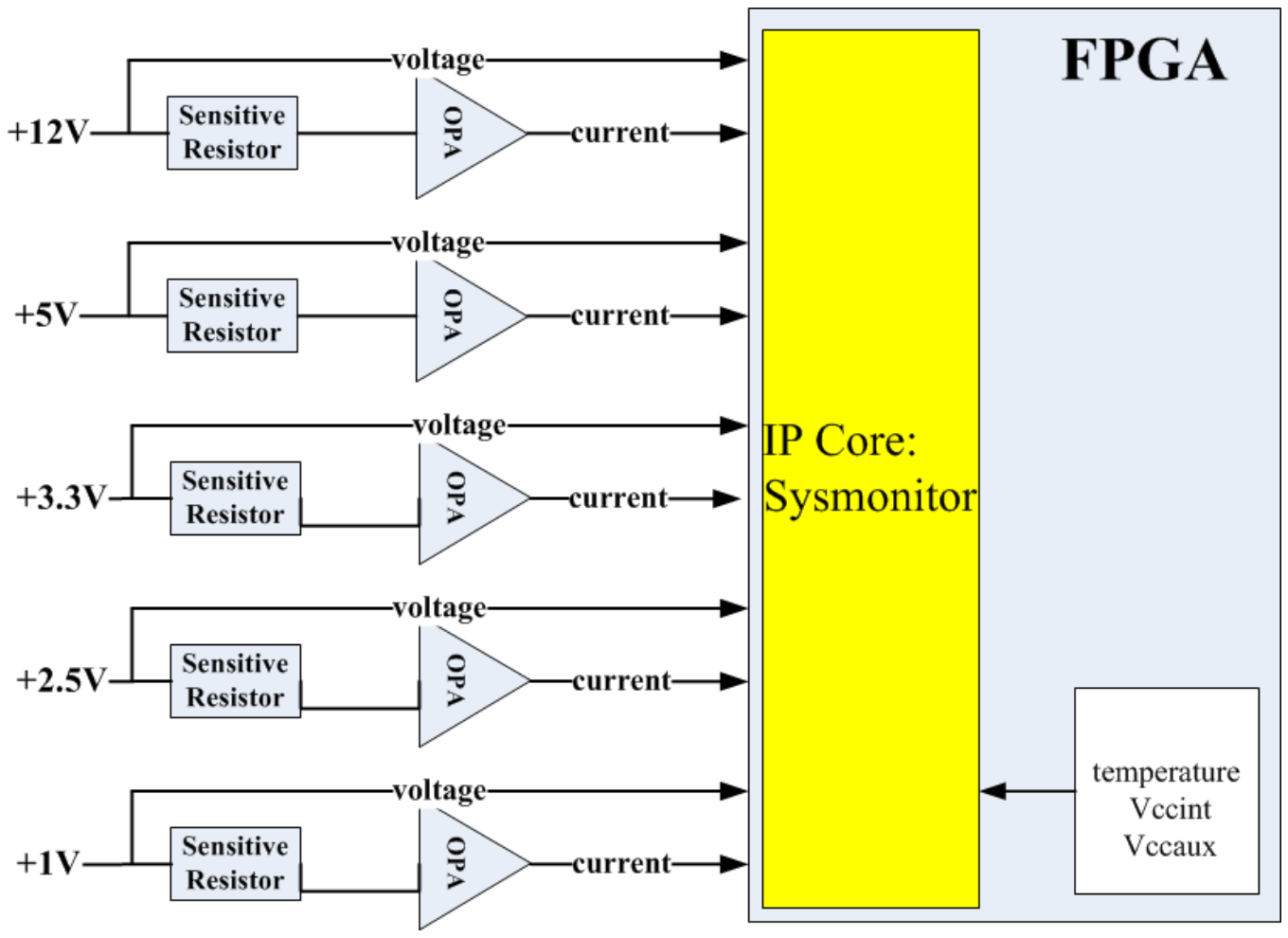

Fig. 8 Block diagram of System monitor module

Taking full advantage of the FPGA's flexibility to program, not only the above functions, but also some necessary modules in the quantum communication experiment such as multi-channel counter and digital-to-analog converter (DAC) control are integrated in a single FPGA.

Besides, 16 channel inputs support separately adjusting the threshold within the range of -2.5V~2.5V, and other 12 low-voltage transistor-transistor logic (LVTTL) channel inputs as a backup. The system also has an external clock input and an external trigger input.

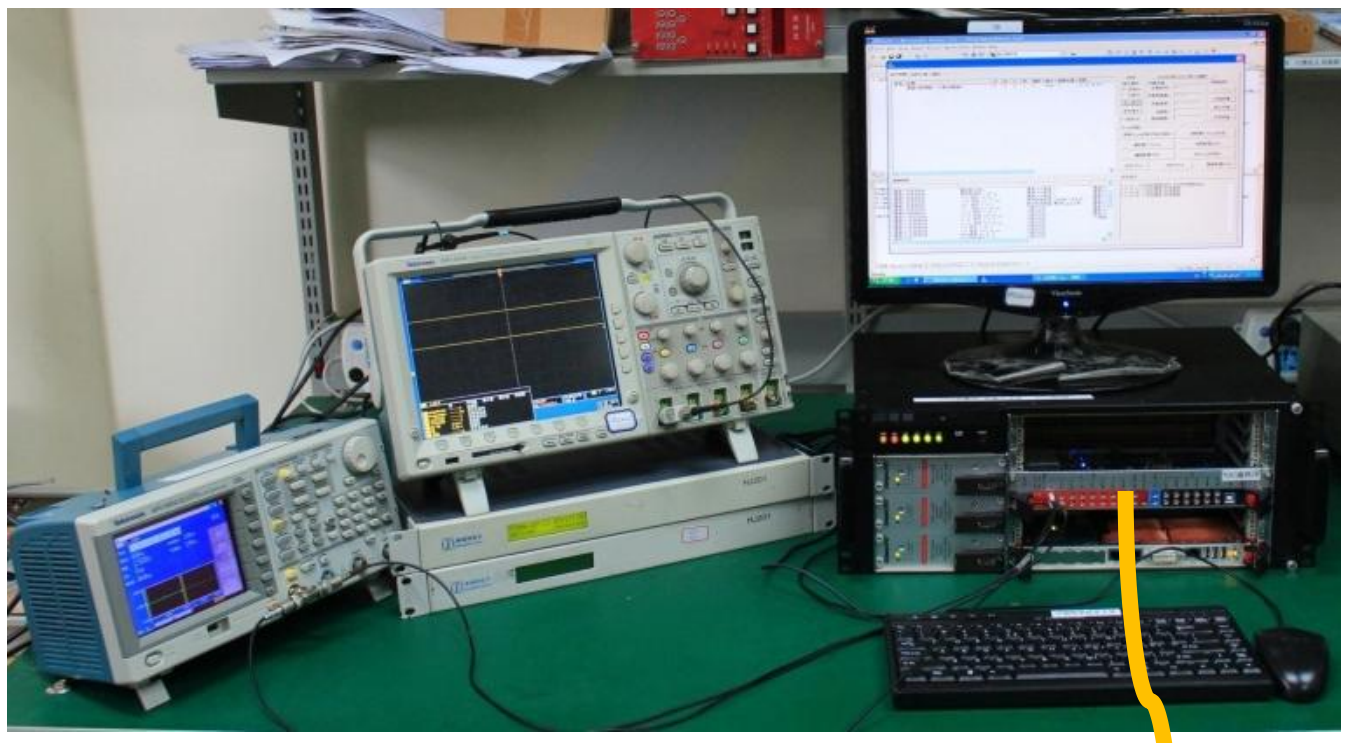

Fig. 9 Electronics performance test setup

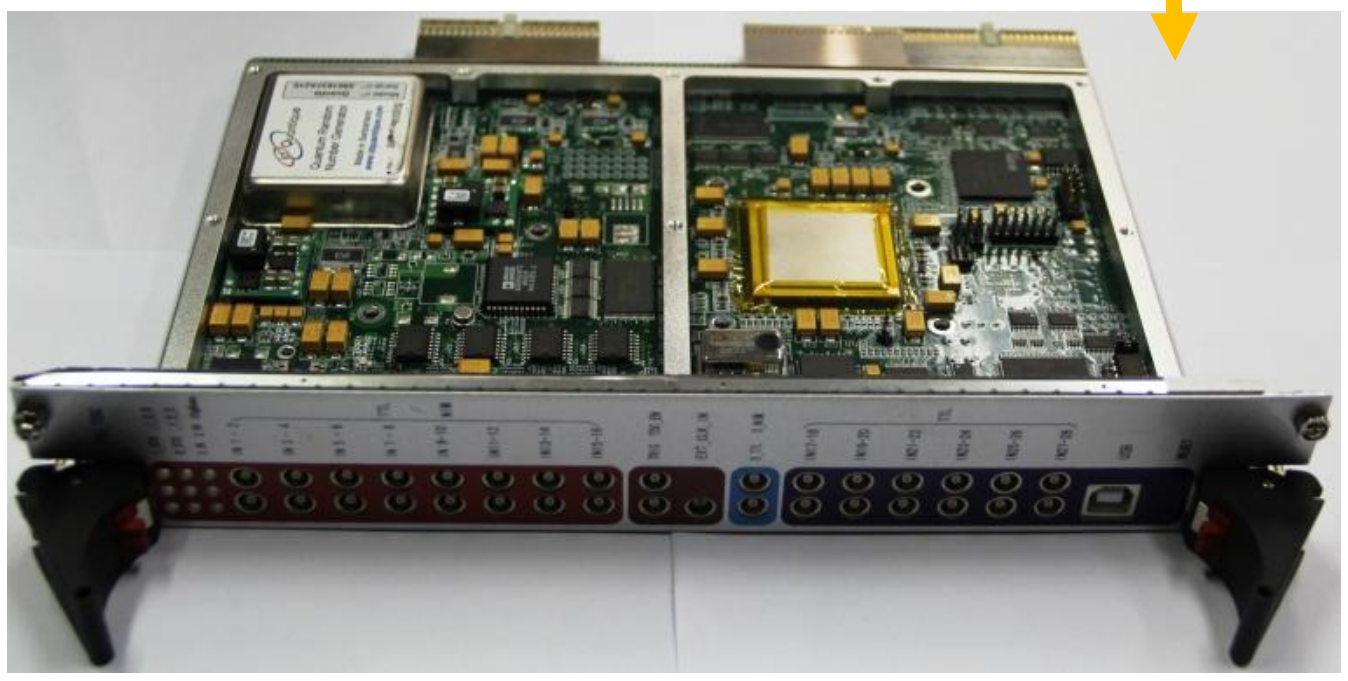

Fig. 10 The electronics board designed on 6U compact PCI

TABLE I. TIME MEASUREMENT PERFORMANCE FOR 16 CHANNELS

| Element | Value |
|---|---|
| RMS | 14ps<RMS<24ps |
| LSB | 23.9ps |
| INL | -3 LSB<INL<3 LSB |
| DNL | -1 LSB<DNL<3 LSB |
| Dead time | 30ns |
| Dynamic range | >1s |

## IV. TEST RESULTS

### A. *Electronics Performance Test*

The electronics performance of this system has been tested. The electronics performance test setup is shown in the Fig. 10, and Fig. 10 shows the standard 6U electronics board designed on CPCI system.

We used cable delays measurements method to evaluate the timing precision of TDC by the period of a hit signal produced by a pulse generator [11]. Fig. 11 shows the typical statistical spread of the time measurements of channel 2 and channel 3, in which the timing precision RMS is about 18.5ps. Fig. 12 and Fig. 13 shows the time non-linearity including DNL and INL. Table I shows the test results of all 16 channels TDC. The time bin size is 23.9ps and the time precision RMS is less than 24ps for 16 channels. The dead time is about 30ns.

The RMS values above are the results of the measured values divided by$\sqrt{2}$, as the measured values of standard deviation contain two TDC channels.

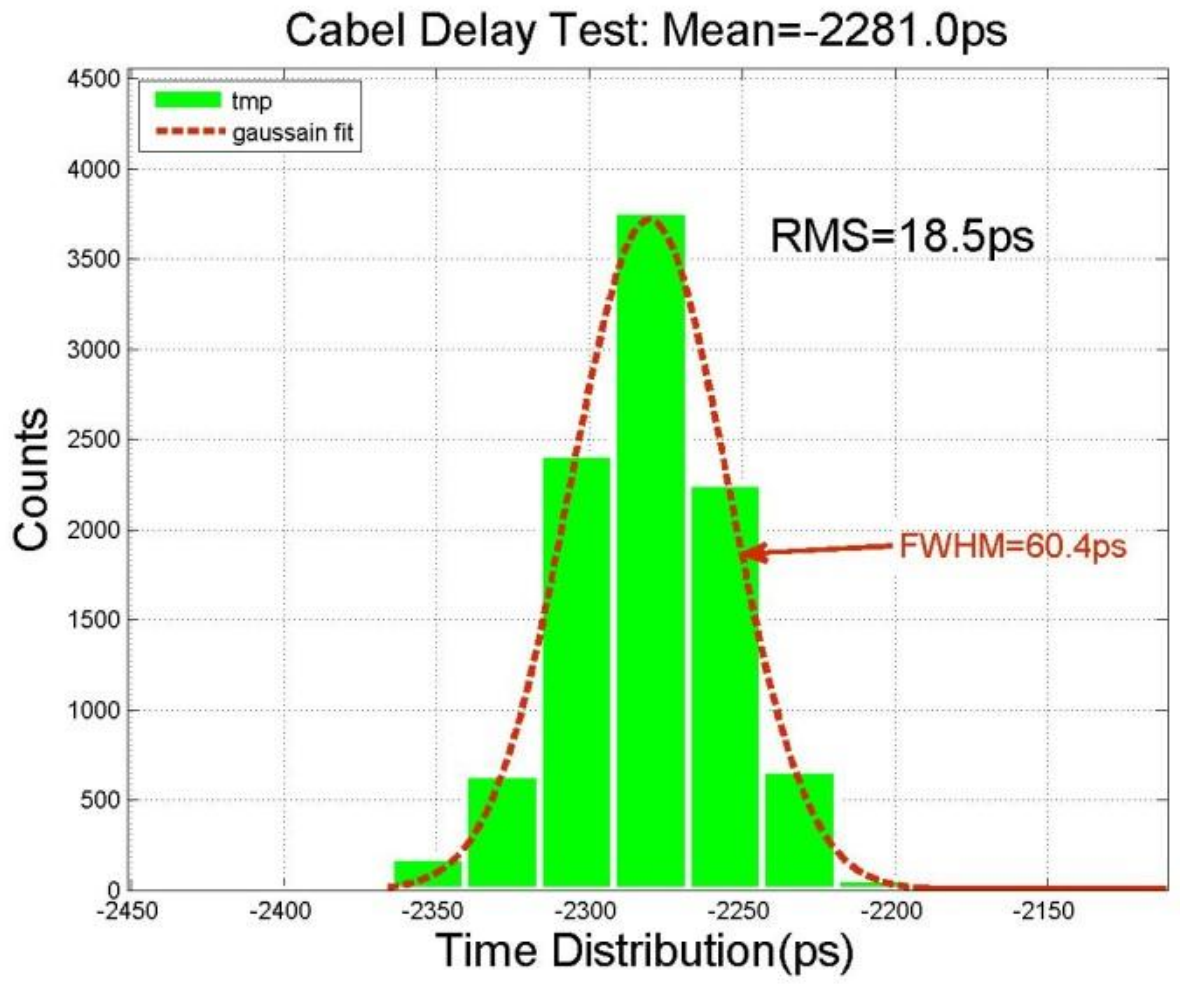

Fig. 11 Typical time precision RMS test result of TDC

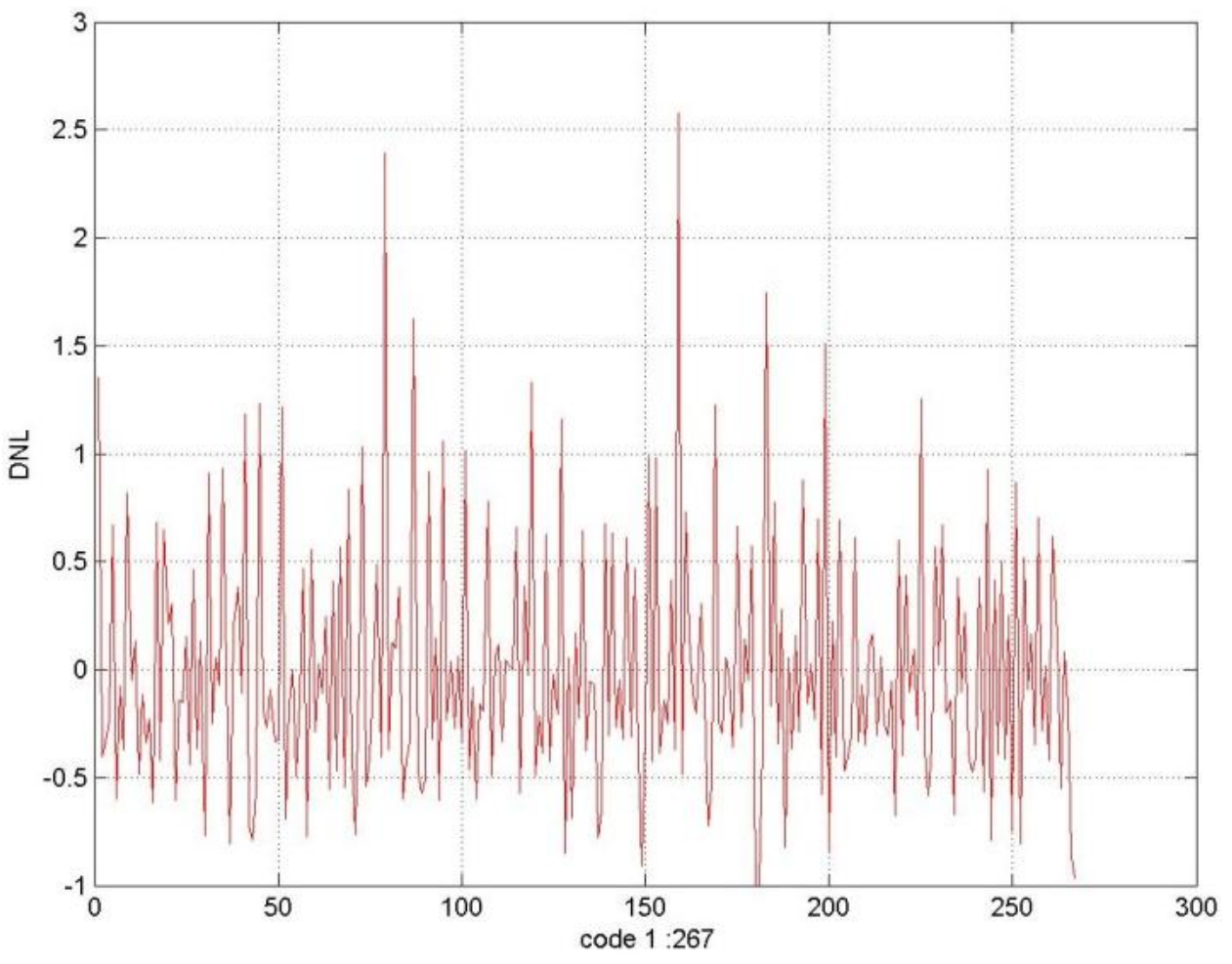

Fig. 12 Typical time DNL performance of TDC

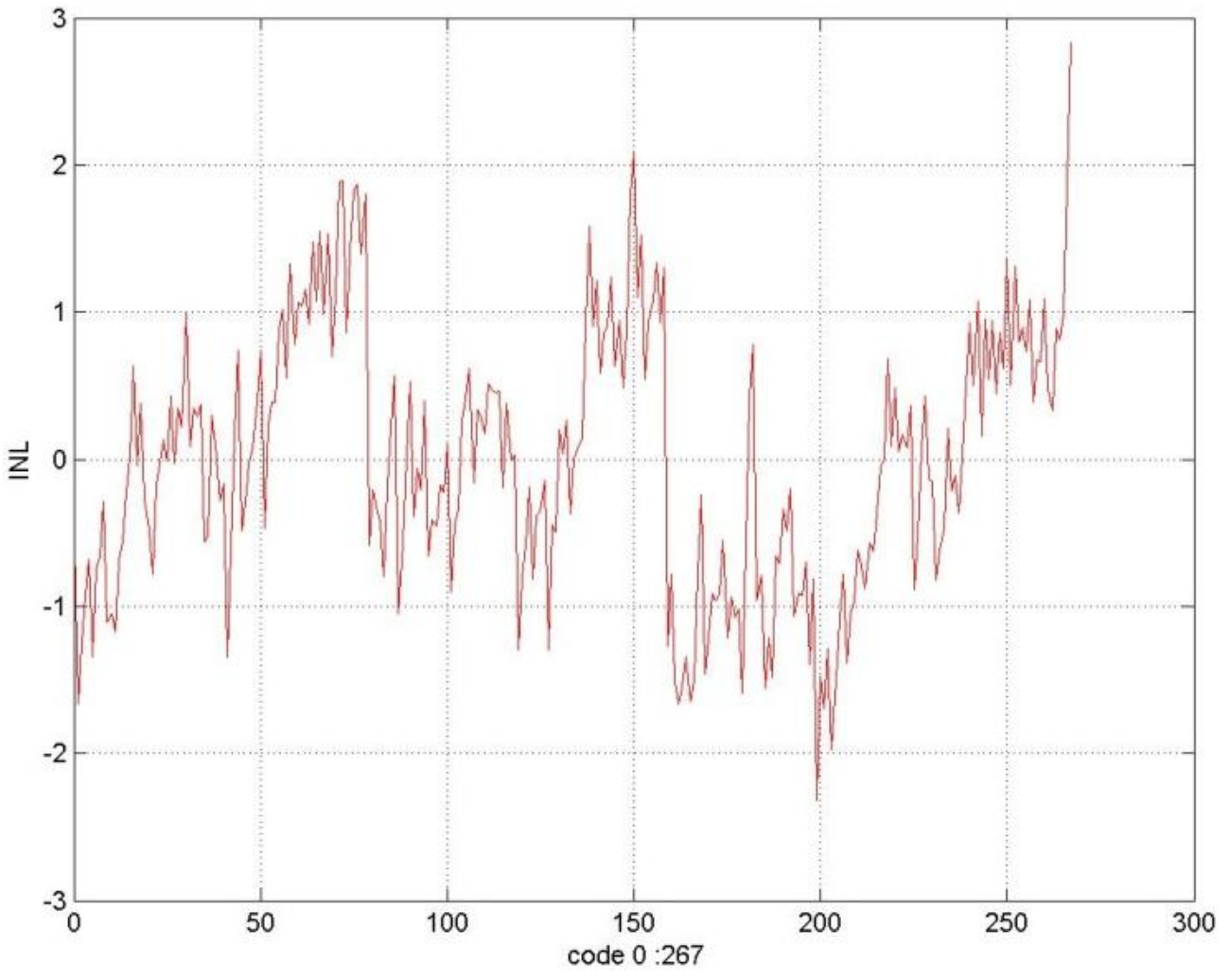

Fig. 13 Typical time INL performance of TDC

The other electronics performance test results are show in Table II. The random number generation rate is up to 4 Mb/s. The data transfer rate to local computer is up to 35 MBps, and the count rate is up to 30M/s. The results show the system can well exactly monitor the operating state of the board, and the Power consumption of the 6U circuit board is less than 9W.

TABLE II. THE OTHER ELECTRONICS PERFORMANCE

| Element | Value |
|---|---|
| Voltage monitoring accuracy | <0.05V |
| Current monitoring accuracy | <20mA |
| Temperature monitoring accuracy | <0.5℃ |
| Max data transfer rate | 35MBps |
| Max count rate | 30M/s |
| Max random bit generating | 4Mb/s |
| Power consumption of 6U board | <9W |

## B. *Performance in QKD Experiment*

Using the experimental set-up shown in Fig. 14, an indoor free space QKD experiment, based on the BB84 protocol, has been carried out to verify the system's performance. This system as Bob communicates with Alice in the experiment. Test results show that the QBER is 1.75% and the final key rate is 1851 bps. It demonstrates that this system can be successfully used and function well in the QKD experiment. What is noteworthy is that the QBER depends on many factors [13] including source error, channel error, detection system error, system time resolution, background count rate, etc., and the final key rate depends on link efficiency, key extraction efficiency, average photon number per signal pulse, QBER, etc.[16].

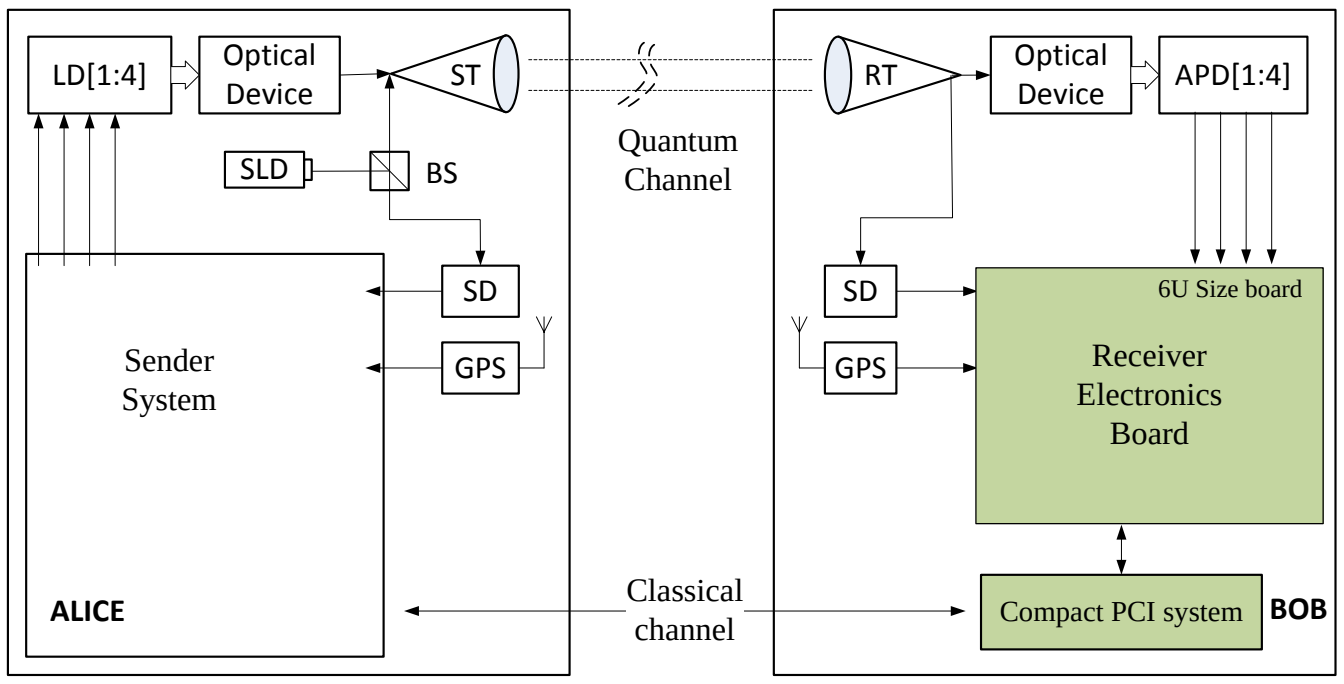

Fig. 14 Block diagram of the free space QKD experiment. SLD, synchronous laser diode; BS, beam splitter; ST, send telescope; RT, receive telescope; SD, synchronous detector; LD[1:4], four laser diodes; APD[1:4], four avalanche photo diodes.

# V. CONCLUSION

A custom and compact electronics system for satellite-ground quantum communication is implemented in this paper, which have a high time measurement precision less than 24 ps for 16 channels and a high data transfer speed up to 35 MBps.

According to the test results, the features of high time resolution, small dead time, high data rate, multi-channel and integratability can well meet the requirements for quantum communication, and it makes the quantum communication electronics more compact and standardized.

All devices used in this system are industrial grade products, which have a better stability and use in more complex environment. The system has been proven to perform well and use successfully in free space QKD experiment. It also can easily be extended and has a wider application in other quantum communication experiments, such as entanglement distribution and quantum teleportation.

# ACKNOWLEDGMENT

The authors would like to thank Chaoze Wang, Peng Shang, Shuangqiang Zhao, Yuhuai Li for their help during the tests and experiment.